\documentclass[aps,twocolumn,showpacs,preprintnumbers,amsmath,amssymb]{revtex4}

\usepackage{graphicx}
\usepackage{dcolumn}
\usepackage{bm}
\usepackage{txfonts}
\usepackage{lscape}
\begin{document}

\title{Four-point vertex in the Hubbard model and partial bosonization}

\author{S. Friederich}  
\author{H. C. Krahl}  
\author{C. Wetterich}

\affiliation{\mbox{\it Institut f{\"u}r Theoretische Physik,
Universit\"at Heidelberg,
Philosophenweg 16, D-69120 Heidelberg, Germany}}

\begin{abstract}
Magnetic and superconducting instabilities in the two-dimensional $t-t'$-Hubbard model are discussed within a functional renormalization group approach. The fermionic four-point vertex is efficiently parametrized by means of partial bosonization. The exchange of composite bosons in the magnetic, charge density and superconducting channels accounts for the increase in the effective couplings with increasing length scale. We compute the pseudocritical temperature for the onset of local order in various channels.
\end{abstract}



\maketitle

\section{Introduction}
The functional renormalization group approach to correlated fermion systems has been of great help to detect different types of instabilities and collective order within many different models. This holds, in particular, for the two-dimensional Hubbard model which is hoped to improve our understanding of superconductivity in the high-$T_c$-cuprates.\cite{zanchi1,zanchi2,halbothmetzner,halbothmetzner2,honisalmi,salmhofer,honerkamp01,katanin} Most studies presented so far rely on the flow of the momentum-dependent four-fermion vertex. (For analogous work on four-quark vertices see \cite{Ellwanger,Meggiolaro}.) They are performed in the so-called $N$-patch scheme where the Fermi surface is discretized into $N$ patches, and the angular dependence of the fermionic four-point function is evaluated only for one momentum on each directional patch.

The approach presented here is based on the introduction of fermionic bilinears corresponding to different types of possible orders (partial bosonization) that was developed and used before in \cite{bbw00,bbw04,bbw05,kw07,krahlmuellerwetterich,simon}. It is also inspired by the efficient parametrization method for the fermionic four-point vertex proposed in \cite{husemann}. The link between the two approaches is given by the fact that different channels of the fermionic four-point function, defined by their (almost) singular momentum structure, correspond to different types of possible orders which are described by different composite bosonic fields.

The advantages of our method are, first, that it allows one to treat the complex momentum dependence of the fermionic four-point function in an efficient, simplified way, involving only a small number of coupled flow equations and second, that it permits to follow the renormalization group flow into phases of broken symmetry. A comparative disadvantage may be a better resolution of contributions from many channels in the $N$-patch approach. (In principle, both approaches can be combined.) In this paper we focus on the first of these two aspects. Spontaneous symmetry breaking was already addressed for antiferromagnetism (AF) \cite{bbw04, bbw05} in the Hubbard model close to half filling and for $d$-wave superconductivity in an effective Hubbard-type model with a dominating coupling in the $d$-wave channel.\cite{kw07}

\begin{figure}[t]
\includegraphics[width=50mm,angle=0.]{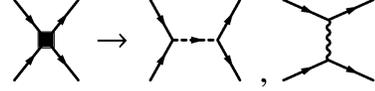}
\caption{\small{Schematic picture of bosonization of the four-fermion vertex. Solid lines correspond to fermions, the dashed line to a complex (Cooper pair) boson, the wiggly line to a real boson representing a particle-hole state in the spin or charge density wave channel.}}
\label{bosonization}
\end{figure}

The main idea behind partial bosonization, namely, to represent the fermionic four-point vertex by a certain number of exchange bosons, is graphically shown in Fig. \ref{bosonization}. The dependence of the four-point vertex on three external frequencies and momenta (or simply ``momenta'', as we are going to write for short) is parametrized in terms of bosonic propagators together with Yukawa couplings which describe the interaction between one boson and two fermions. The different bosonic channels are distinguished according to the structure of their momentum dependence which may possibly become singular due to a zero of the inverse boson propagator. The momentum-independent part of the four-fermion vertex may either be distributed onto the different bosonic channels or be kept fixed as a purely fermionic coupling. In order to avoid the arbitrariness encountered when one chooses the first of these two options, we adopt the second for our computations. This may be regarded as a prototype for a combination of partial bosonization with the $N$-patch method in the sense of keeping only one patch and setting the four-fermion coupling to a constant value.

Although for a numerically exact treatment of the four-fermion vertex an infinity of bosonic fields would have to be considered in principle, a small number of well-chosen fields may suffice for a reasonable quantitative precision. The choice of fields that has to be included in order to capture the relevant physics depends on the model under investigation. In the case of the two-dimensional Hubbard model at small next-to-nearest-neighbor hopping $|t'|$, a magnetic boson $\mathbf{m}$ and a $d$-wave Cooper pair boson $d$ are needed because they correspond to the instabilities that occur. In order to avoid a too poor momentum resolution of the four-fermion vertex, we also include an $s$-wave Cooper pair boson field $s$ and a charge density boson field $\rho$. Other types of bosons are needed in other contexts---for instance, a $d$-wave charge density boson for the study of Pomeranchuk instabilities or a $p$-wave boson for triplett superconductivity at larger values of $|t'|$ away from the van Hove filling \cite{honisalmi}. With the restriction to the bosons $\mathbf{m},d,s$ and $\rho$, supplemented by a pointlike four-fermion vertex, we will show that interesting results and a semiquantitative understanding can be found in a rather simple truncation.

The two-dimensional Hubbard model \cite{hubbard,kanamori,gutzwiller} on a square lattice has attracted a lot of attention in the past 25 years because it is thought to cover important aspects of the physics of the high-$T_c$ cuprates. In analogy to the phase diagram of the cuprates, it shows antiferromagnetic order at half-filling and is believed to exhibit $d$-wave superconducting order (dSC) away from half filling.\cite{anderson} Today there are many studies which predict the $d$-wave instability to be the dominating one in a certain range of parameters aside from half filling \cite{miyake,loh,bickers,lee,millis,monthoux,scalapino,bickersscalapinowhite,bulut,pruschke,maier,senechal,maierjarrellscalapino,maiermacridin}, for a systematic overview see \cite{scalapinoreview}. The picture is also confirmed by some strikingly simple scaling approaches \cite{schulz,dzyaloshinskii,lederer} and finds further support within more elaborate renormalization group studies such as \cite{zanchi1,zanchi2,halbothmetzner,halbothmetzner2,honisalmi,salmhofer,honerkamp01,metznerreissrohe,reiss}.

\section{Method and approximation}
The starting point of our treatment is the exact flow equation for the effective average action or flowing action,\cite{cw93}
\begin{equation} \label{floweq}
\partial_k \Gamma_k = \frac{1}{2} \rm{STr} \,
  \left(\Gamma^{(2)}_k + R_k\right)^{-1}  \partial_k R_k =
 \frac{1}{2} \rm{STr}\,\tilde \partial_k \,\left(\ln  (\Gamma^{(2)}_k + R_k)\right)\,.
\end{equation}
The dependence on the renormalization group scale $k$ is introduced by adding a regulator $R_k$ to the full inverse propagator $\Gamma^{(2)}_k$. In Eq. (\ref{floweq}), $\rm{STr}$ denotes a supertrace, which sums over momenta, frequencies, and internal indices, while $\tilde \partial_k$ is the scale derivative acting only on the infrared (IR) regulator $R_k$. The Hamiltonian of the system under considerations is taken into account by the initial condition $\Gamma_{k=\Lambda}=S$ of the renormalization flow, where $\Lambda$ denotes some very large UV scale and $S$ is the microscopic action in a functional integral formulation of the Hubbard model. In the IR limit ($k\to 0$) the flowing action $\Gamma_k$ equals the full effective action $\Gamma=\Gamma_{k\to 0}$, which is the generating functional of one-particle-irreducible (1PI) vertex functions.

We employ a compact notation with $Q=(\omega_n=2\pi nT,\mathbf{q})$ and $Q=(\omega_n=(2n+1)\pi T,\mathbf{q})$ for bosonic and fermionic fields and
\begin{eqnarray}\label{eq:sumdefinition}
&&\quad\sum\limits_Q=T\sum\limits_{n=-\infty}^\infty \int\limits_{-\pi}^\pi \frac{d^2q}{(2\pi)^2}\,,\nonumber\\
&&\delta(Q-Q')=T^{-1}\delta_{n,n'}(2\pi)^2\delta^{(2)}(\mathbf{q}-\mathbf{q'})\,.
\end{eqnarray}
The components of the momentum $\mathbf q$ are measured in units of the inverse lattice distance $\mathrm{a}^{-1}$. The discreteness of the lattice is reflected by the $2\pi$-periodicity of the momenta $\mathbf{q}$.

Although Eq. \eqref{floweq} is an exact flow equation, it can only be solved approximately. In particular, a truncation has to be specified for the flowing action, indicating which of the (infinitely many) 1PI vertex functions are actually taken into account. Our ansatz for the flowing action includes contributions for the electrons, for the bosons in the magnetic, charge, and $s$-wave and $d$-wave superconducting channels, and for interactions between fermions and bosons,
\begin{eqnarray}
\Gamma_k[\chi]&=&\Gamma_{F,k}[\chi]+\Gamma_{Fm,k}[\chi]+\Gamma_{F\rho,k}[\chi]+\Gamma_{Fs,k}[\chi]+\Gamma_{Fd,k}[\chi]\nonumber\\
       &&+\Gamma_{m,k}[\chi]+\Gamma_{\rho,k}[\chi]+\Gamma_{s,k}[\chi]+\Gamma_{d,k}[\chi]\,.
\end{eqnarray}
The collective field $\chi=(\mathbf m,\rho,s,s^*,d,d^*,\psi,\psi*)$ includes both fermion fields $\psi,\psi*$ and boson fields $\mathbf m,\rho,s,s^*,d,d^*$.

The purely fermionic part $\Gamma_{F}[\chi]$ (the dependence on the scale $k$ is always implicit in what follows) of the flowing action consists of a two-fermion kinetic term $\Gamma_{F\rm{kin}}$, a momentum-independent four-fermion term $\Gamma_F^U$, and the momentum-dependent four-fermion terms $\Gamma_F^m$, $\Gamma_F^\rho$, $\Gamma_F^s$ and $\Gamma_F^d$,
\begin{eqnarray}
\Gamma_{F}[\chi]=\Gamma_{F\rm{kin}}+\Gamma_{F}^U+\Gamma_F^m+\Gamma_F^\rho+\Gamma_F^s+\Gamma_F^d\,.
\end{eqnarray}
The fermionic kinetic term is given by
\begin{eqnarray}\label{fermprop}
\Gamma_{F\rm{kin}}=\sum_{Q}\psi^{\dagger}(Q)P_F(Q)\psi(Q)\,,
\end{eqnarray}
with inverse fermion propagator
\begin{eqnarray}\label{PF}
P_{F}(Q)=i\omega+\xi(\mathbf q)
\,,\end{eqnarray}
where we take for the dispersion relation of the free electrons
\begin{equation}
\xi(\mathbf q)=-\mu-2t(\cos q_x +\cos q_y)-4t' \cos q_x\cos q_y\,.
\end{equation}

The momentum-independent part of the four-fermion coupling is identical to the Hubbard interaction $U$. In our truncation, this coupling is not modified during the flow. The corresponding part of the effective action therefore reads
\begin{eqnarray}
\Gamma_{F}^U &=&\frac{1}{2}\sum_{K_1,K_2,K_3,K_4}U\,\delta\left( K_1-K_2+K_3-K_4 \right)\,\nonumber\\&&\hspace{0.5cm}\times\,\big\lbrack\psi^\dagger(K_1)\psi(K_2)\big\rbrack\,\big\lbrack\psi^\dagger(K_3)\psi(K_4)\big\rbrack\,.
\end{eqnarray}

In this work, as in \cite{zanchi1,zanchi2,halbothmetzner,halbothmetzner2,honisalmi,salmhofer,honerkamp01,katanin}, contributions to the fermionic self-energy are neglected. Instead, we focus on the momentum dependence of the fermionic four-point function $\lambda_F(K_1,K_2,K_3,K_4)$, which, due to energy-momentum conservation $K_4=K_1-K_2+K_3$, is a function of three independent momenta. We decompose this vertex into a sum of four functions $\lambda_F^m(Q)$, $\lambda_F^\rho(Q)$, $\lambda_F^s(Q)$ and $\lambda_F^d(Q)$, each depending on only one particular combination of the $K_i$, which correspond to the four different bosons taken into account. This is inspired by the singular momentum structure of the leading contributions to the four-fermion vertex. In our ansatz for the effective average action these functions enter as
\begin{align}
\Gamma_F^m=-\frac{1}{2}\sum_{K_1,K_2,K_3,K_4}\lambda_F^m(K_1-K_2)\,\delta\left( K_1-K_2+K_3-K_4 \right)\nonumber\\
\times\,\big\lbrack \psi^\dagger(K_1)\boldsymbol\sigma\psi(K_2) \big\rbrack\cdot\big\lbrack \psi^\dagger(K_3)\boldsymbol\sigma\psi(K_4) \big\rbrack\,,\label{aform}\\
\Gamma_F^\rho=-\frac{1}{2}\sum_{K_1,K_2,K_3,K_4}\lambda_F^\rho(K_1-K_2)\,\delta\left( K_1-K_2+K_3-K_4 \right)\nonumber\\
\times\,\big\lbrack \psi^\dagger(K_1)\psi(K_2) \big\rbrack\,\big\lbrack \psi^\dagger(K_3)\psi(K_4) \big\rbrack
\end{align}
for the real bosons, and, for the superconducting bosons, as
\begin{align}
\Gamma_F^s=\sum_{K_1,K_2,K_3,K_4}\lambda_F^s(K_1+K_3)\,\delta\left( K_1-K_2+K_3-K_4 \right)\nonumber\\
\times\,\big\lbrack \psi^\dagger(K_1)\epsilon\psi^*(K_3) \big\rbrack\,\big\lbrack \psi^T(K_2)\epsilon\psi(K_4) \big\rbrack\,,\\
\Gamma_F^d=\sum_{K_1,K_2,K_3,K_4}\lambda_F^d(K_1+K_3)\,\delta\left( K_1-K_2+K_3-K_4 \right)\nonumber\\
\times\, f_d(( K_1-K_3)/2)\, f_d(( K_2-K_4)/2)\nonumber\\
\times\,\big\lbrack \psi^\dagger(K_1)\epsilon\psi^*(K_3) \big\rbrack\,\big\lbrack \psi^T(K_2)\epsilon\psi(K_4) \big\rbrack\,,\label{dform}
\end{align}
where $\boldsymbol\sigma=(\sigma^1,\sigma^2,\sigma^3)$ is the vector of the Pauli matrices, the matrix $\epsilon$ is defined as $\epsilon=i\sigma^2$, and the function
\begin{equation}
f_d(Q)=f_d(\mathbf q)=\frac{1}{2}\left( \cos{q_x}-\cos{q_y} \right)\,
\end{equation}
is the $d$-wave form factor which is kept fixed during the flow.

In a first step, contributions to the four-fermion vertex are distributed onto the couplings $\lambda_F^m,\;\lambda_F^\rho,\;\lambda_F^s,\;\lambda_F^d$, depending on their momentum dependence. Partial bosonization comes into play at this stage as the absorption of these contributions by the corresponding Yukawa couplings and bosonic propagators. More concretely, this means that the couplings $\lambda_F^m,\;\lambda_F^\rho,\;\lambda_F^s,\;\lambda_F^d$ are set to zero by introducing a scale-dependence of the bosonic fields, which in turn generates additional contributions to the various Yukawa couplings. The technique by means of which this is achieved is called flowing bosonization or rebosonization.\cite{GiesWett,floerchi} We describe it in some detail in Appendix A. In consequence, the complicated spin and momentum dependence of the fermionic four-point function $\lambda_F(K_1,K_2,K_3,K_4)$, as it emerges during the flow, will be captured by the momentum dependence of the propagators of the bosons and the couplings between bosons and fermions.

The interaction between electrons and composite bosons are taken into account in our ansatz for the flowing action by Yukawa-type vertices of the form
\begin{eqnarray}\label{GFak}
\Gamma_{Fm}&=&-\!\sum_{K,Q,Q'}\bar h_m(K)\;\mathbf{m}(K)\cdot[\psi^{\dagger}(Q)\boldsymbol{\sigma}\psi(Q')]\;\delta(K-Q+Q')\,,\nonumber\\
\Gamma_{F\rho}&=&-\!\sum_{K,Q,Q'}\bar h_\rho(K)\;\rho(K)\,[\psi^{\dagger}(Q)\psi(Q')]\;\delta(K-Q+Q')\,,\nonumber\\
\Gamma_{Fs}&=&-\!\sum_{K,Q,Q'}\bar h_s(K)\,\left(s^*(K)\,[\psi^{T}(Q)\epsilon\psi(Q')]\right.\\
&&\hspace{1.5cm}\left.-s(K)\,[\psi^{\dagger}(Q)\epsilon\psi^*(Q')]\right)\;\delta(K-Q-Q')\,,\nonumber\\
\Gamma_{Fd}&=&-\!\sum_{K,Q,Q'}\bar h_d(K)f_d\,\left((Q-Q')/2\right)\left(d^*(K)\,[\psi^{T}(Q)\epsilon\psi(Q')]\right.\nonumber\\
&&\hspace{1.5cm}\left.-d(K)\,[\psi^{\dagger}(Q)\epsilon\psi^*(Q')]\right)\;\delta(K-Q-Q')\,.\nonumber
\end{eqnarray}
Note the presence of the $d$-wave form factor in the second-to-last line. To determine the $k$-dependence of the Yukawa couplings $\bar h_m, \bar h_\rho, \bar h_s, \bar h_d$ is a central task within our approach.

The purely bosonic parts of the effective action are characterized by the bosonic propagators. For the magnetic boson, for instance, the inverse propagator is given by $\tilde P_{m}(Q)\equiv P_{m}(Q)+\bar m_m^2$, where $\bar m_m^2$ is its minimal value and $P_{m}(Q)$ is the (strictly positive) so-called kinetic term. The contributions to the effective average action where the bosonic propagators appear are
\begin{eqnarray}
\Gamma_{m}&=&\frac{1}{2}\sum_{Q}\mathbf{m}^{T}(-Q)\left(P_{m}(Q)+\bar m_m^2\right)\mathbf{m}(Q)\,,\label{m_masse}\\
\Gamma_{\rho}&=&\frac{1}{2}\sum_{Q}\rho(-Q)\left(P_{\rho}(Q)+\bar m_\rho^2\right)\rho(Q)\,,\\
\Gamma_{s}&=&\sum_{Q}s^*(Q)\left(P_{s}(Q)+\bar m_s^2\right)s(Q)\,,\\
\Gamma_{d}&=&\sum_{Q}d^*(Q)\left(P_{d}(Q)+\bar m_d^2\right)d(Q)\,.
\end{eqnarray}
Our parametrization of the frequency and momentum dependence of the bosonic propagators and the Yukawa couplings is described in Appendix B. In contrast to the decomposition of the fermionic four-point vertex proposed in \cite{husemann}, our bosonic propagators exhibit an explicit frequency dependence.

In the present paper, the purely bosonic parts of the flowing action are confined to the bosonic propagators. Higher order purely bosonic interactions are currently investigated and will be included in a forthcoming work.

\section{Initial Conditions and Regulators}

At the microscopic scale $k=\Lambda$ the flowing action must be equivalent to the microscopic action of the Hubbard model, so the initial value of the four-fermion coupling must correspond to the Hubbard interaction $U$. The bosonic fields decouple completely at this scale, so the initial values of the Yukawa couplings are
\begin{equation}
\bar h_m|_\Lambda=\bar h_\rho|_\Lambda=\bar h_s|_\Lambda=\bar h_d|_\Lambda=0\,.
\end{equation}
The purely bosonic part of the effective action on initial scale is set to
\begin{eqnarray}\label{eq:initialcond}
\Gamma_{m}|_{\Lambda}=\mathbf m^T\cdot\mathbf m\,,\quad \Gamma_{\rho}|_{\Lambda}=\rho^T\rho\,,\\
\Gamma_{s}|_{\Lambda}=s^*s\,,\quad \Gamma_{d}|_{\Lambda}=d^*d\,.\nonumber
\end{eqnarray}
In other words, we take $\bar m_{i,\Lambda}^2=t^2$ and then use units $t=1$ and $P_{i,\Lambda}=0$. The choice $\bar m_{i,\Lambda}^2=t^2$ amounts to an arbitrary choice for the normalization of the bosonic fields, which are introduced as redundant auxiliary fields at the scale $\Lambda$, where they do not couple to the electrons. Of course, this changes during the flow, where the bosons are transformed into dynamical composite degrees of freedom, with nonzero Yukawa couplings  and a nontrivial momentum dependence of their propagators.

In addition to the truncation of the effective average action, regulator functions for both fermions and bosons have to be specified. We use ``optimized cutoffs'' \cite{litim1,litim2} for both fermions and bosons. The regulator function for fermions is given by
\begin{eqnarray}
R^F_k(Q)=\rm{sgn}(\xi(\mathbf q))\left(k-|\xi(\mathbf q)|\right)\Theta(k-|\xi(\mathbf q)|)\,,
\end{eqnarray}
the regulator functions for the real bosons are given by
\begin{eqnarray}\label{regulator}
R^{m/\rho}_k(Q)=A_{m/\rho}\cdot(k^2-F_{c/i}(\mathbf q,\hat q))\Theta(k^2-F_{c/i}(\mathbf q,\hat q))
\,\end{eqnarray}
allowing for an incommensurability $\hat q$ with $F_{c/i}$ as defined in Appendix B. Regulator functions for the Cooper-pair bosons are of the same form, but no incommensurability needs to be accounted for in these cases.

\section{Flow equations}

\begin{figure}[t]
\includegraphics[width=45mm,angle=0.]{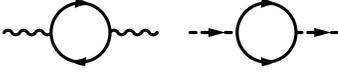}
\caption{\small{1PI diagrams contributing to the flow of bosonic propagators. Wiggly lines denote real bosons (particle-hole channels), dashed lines complex bosons (Cooper pair channels).}}
\label{bosepropkorr}
\end{figure}

The flow equations for the couplings follow from projection of the flow equation for the flowing action onto the various different monomials of fields. The right-hand sides of these flow equations are given by the one-particle-irreducible diagrams having an appropriate number of external lines, including a scale derivative $\tilde \partial_k$ acting only on the IR regulator $R_k$. Diagrams contributing to the flow of boson propagators are shown in Fig. \ref{bosepropkorr}.

Once some bosonic mass term $\bar m_i^2$ changes sign from positive to negative during the flow, this signals the divergence of the four-fermion vertex function in the corresponding channel. A negative mass term indicates local order, since at a given coarse graining scale $k$ the effective average action evaluated at constant field has a minimum for a nonzero value of the boson field. The largest temperature where at fixed values of $U,t',\mu$ one of the mass terms $\bar m_i^2$ changes sign during the flow is called the pseudocritical temperature $T_{pc}$. It can also be described as the largest temperature where short-range order sets it. If this order persists for $k$ reaching a macroscopic scale, the model exhibits effectively spontaneous symmetry breaking, associated in our model to (either commensurate or incommensurate) antiferromagnetism or $d$-wave superconductivity. The largest temperature for which local order persists up to some $k$ corresponding to the inverse size of a macroscopic sample is the true critical temperature $T_c$. In this paper we focus on the symmetric regime where we have a positive mass term and stop the flow once a mass term reaches zero. We plan to address the symmetry-broken regimes in a future work.

\begin{figure}[t]
\includegraphics[width=50mm,angle=0.]{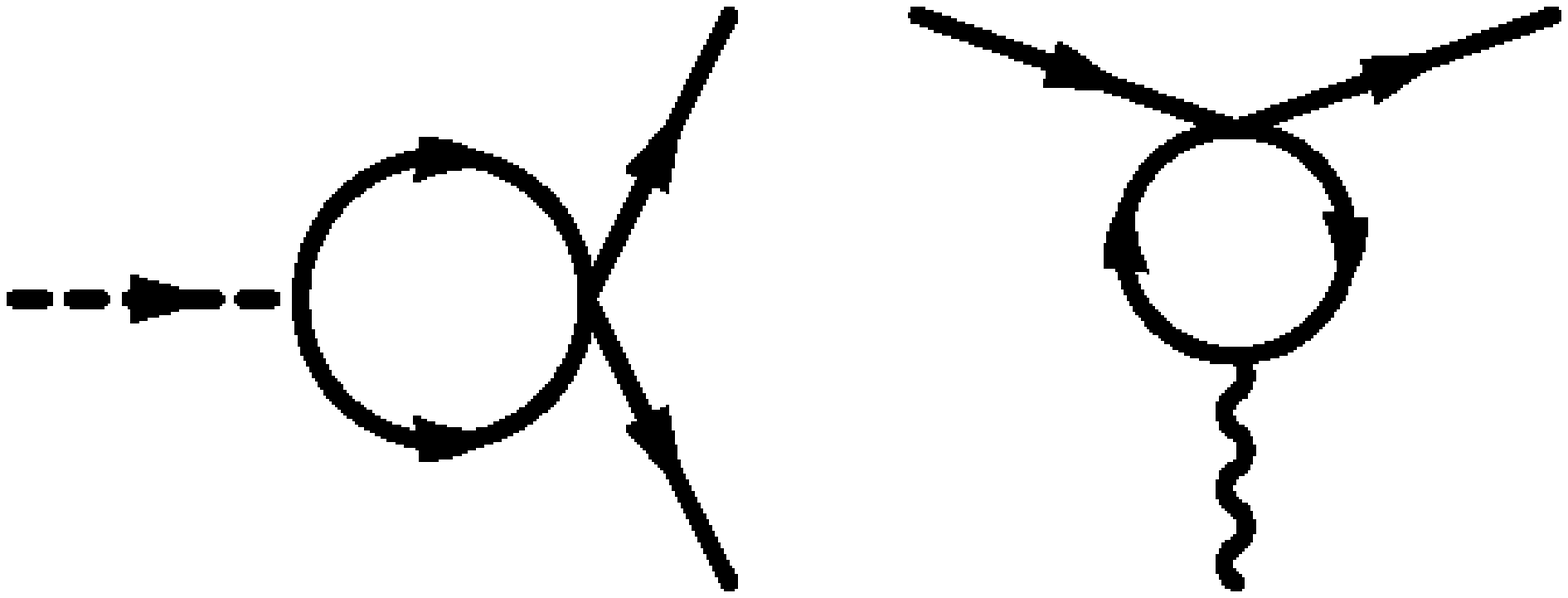}\vspace{0.5cm}
\includegraphics[width=65mm,angle=0.]{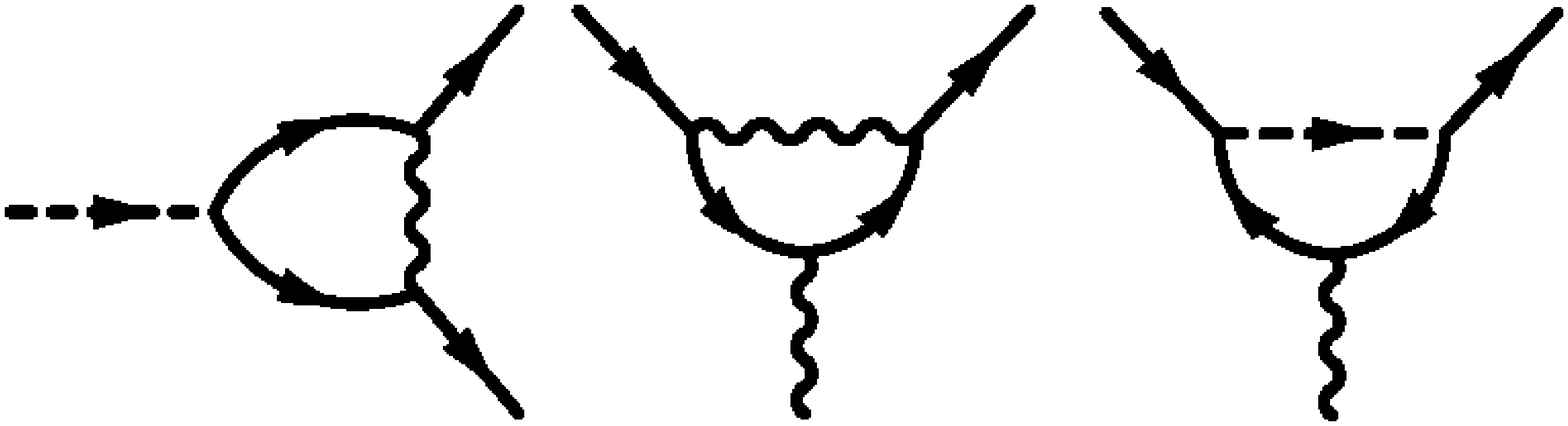}
\caption{\small{1PI diagrams which directly contribute to the flow of the Yukawa couplings.}}
\label{yukdir}
\end{figure}

\begin{figure}[t]
\includegraphics[width=40mm,angle=0.]{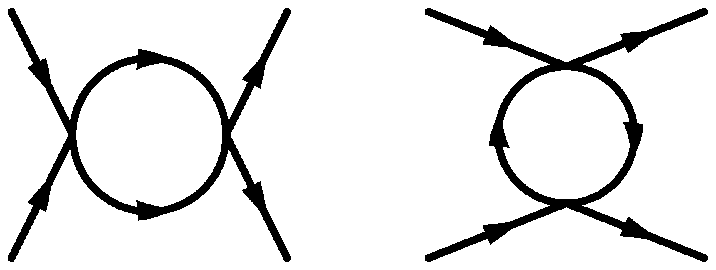}\vspace{0.5cm}
\includegraphics[width=60mm,angle=0.]{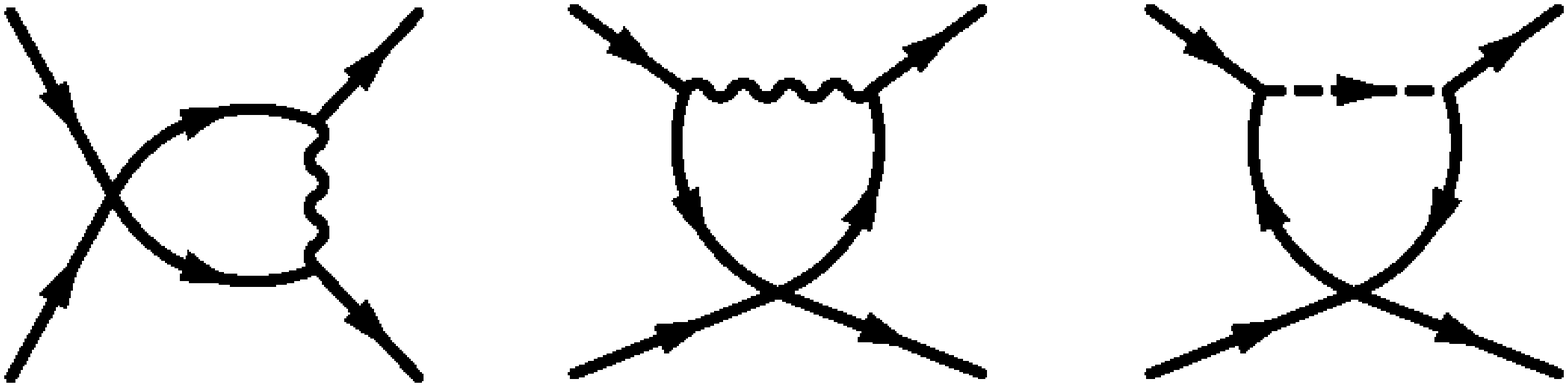}\vspace{0.5cm}
\includegraphics[width=80mm,angle=0.]{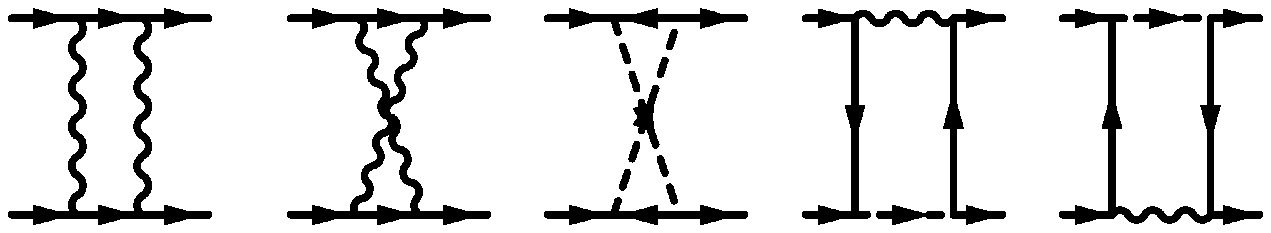}
\caption{\small{1PI diagrams contributing to the flow of the Yukawa couplings via flowing bosonization.}}
\label{ff}
\end{figure}

The flow equations for the Yukawa couplings consist of a direct contribution and an ``indirect'' contribution resulting from flowing bosonization, see Appendix A. Diagrams contributing directly to the flow of the Yukawa couplings are shown in Fig. \ref{yukdir}, those that contribute via flowing bosonization are displayed in Fig. \ref{ff}. Since we choose to distribute contributions from flowing bosonization only onto the Yukawa couplings and not onto the masses, it is crucial to include a momentum dependence of the Yukawa coupling $\bar h_m$ in the magnetic channel in order to account for the emergence of the $d$-wave superconducting instability. Otherwise the contribution of the particle-particle box diagram (the first in the lower line of Fig. \ref{ff}) to the $d$-wave coupling would be underestimated.

In order to demonstrate how the contributions to the four-fermion vertex are taken into account via flowing bosonization, we discuss the case of the purely fermionic loop diagrams shown in the upper line of Fig. \ref{ff}. As long as no scale dependence due to the regulator function has been introduced, they are given by
\begin{eqnarray}
\Delta\Gamma^F_F&=&-\frac{U^2}{2}\sum_{K_1,K_2,K_3,K_4}\sum_P \\
&&\hspace{-0.6cm}\left( \frac{1}{P_F(P)P_F(P+K_2-K_3)} + \frac{1}{P_F(P)P_F(-P+K_1+K_3)}\right) \nonumber\\
&&\hspace{-0.6cm}\delta\left( K_1-K_2+K_3-K_4 \right)\big\lbrack \psi^\dagger(K_1)\psi(K_2) \big\rbrack\cdot\big\lbrack \psi^\dagger(K_3)\psi(K_4) \big\rbrack\,.\nonumber
\end{eqnarray}
In order to obtain the resulting contribution to the fermionic four-point vertex function $\Delta\Gamma^{F\,(4)}_F$, we have to take the fourth functional derivative of $\Delta\Gamma^F_F$ with respect to the fermionic fields. It is given by
\begin{eqnarray}
&&\Delta\Gamma^{F\,(4)}_F(K_1,K_2,K_3,K_4)=\label{oneloop}\\
&&\frac{1}{4}\frac{\delta^4}{\delta\psi^*_\alpha(K_1)\delta\psi_\beta(K_2)\delta\psi^*_\gamma(K_3)\delta\psi_\delta(K_4)}\Delta\Gamma_F^F \nonumber\\
&&=-\frac{U^2}{4}\sum_P\Big\lbrace \frac{4\,S_{\alpha\gamma;\beta\delta}}{P_F(P)P_F(-P+K_1+K_3)}\nonumber\\
&&- \frac{\delta_{\alpha\delta}\delta_{\gamma\beta}}{P_F(P)P_F(P+K_2-K_1)}
+ \frac{\delta_{\alpha\beta}\delta_{\gamma\delta}}{P_F(P)P_F(P+K_2-K_3)}
\Big\rbrace\nonumber\,,
\end{eqnarray}
where $S_{\alpha\gamma;\beta\delta}=\frac{1}{2}\left( \delta_{\alpha\beta}\delta_{\gamma\delta} - \delta_{\alpha\delta}\delta_{\gamma\beta} \right)$ denotes the singlet projection. The two last lines of Eq. \eqref{oneloop} can be compared to the fourth derivative with respect to the fields of the right hand sides of Eqs. \eqref{aform}\,-\,\eqref{dform}. This allows one to obtain the loop corrections to the four-fermion couplings $\lambda_F^m,\;\lambda_F^\rho,\;\lambda_F^s,\;\lambda_F^d$ introduced there. The second last line of Eq. \eqref{oneloop} can be absorbed by the $s$-boson, the last line by the $\mathbf m$- and $\rho$-bosons. No contribution to the $d$-boson arises at this stage.

To determine how the last line of Eq. \eqref{oneloop} should be distributed onto the $\mathbf m$- and $\rho$-bosonic channels, we use the identity $\delta_{\alpha\delta}\delta_{\gamma\beta}=\frac{1}{2}\left( \delta_{\alpha\beta}\delta_{\gamma\delta}+\sigma^j_{\alpha\beta}\sigma^j_{\gamma\delta}  \right)$\,. All terms have now the same structures as those appearing in the fourth functional derivative of Eq. \eqref{aform}. We obtain the following loop contributions to $\lambda_F^m,\;\lambda_F^\rho,\;\lambda_F^s$:
\begin{eqnarray}
\left(\Delta\lambda_F^m\right)^F(K_1-K_2)=-\frac{U^2}{2}\sum_P\frac{1}{P_F(P)P_F(P+K_2-K_1)}\,,\nonumber\\
\left(\Delta\lambda_F^\rho\right)^F(K_1-K_2)=-\frac{U^2}{2}\sum_P\frac{1}{P_F(P)P_F(P+K_2-K_1)}\,,\label{simpleloop}\\
\left(\Delta\lambda_F^s\right)^F(K_1+K_3)=\frac{U^2}{2}\sum_P\frac{1}{P_F(P)P_F(-P+K_1+K_3)}.\nonumber
\end{eqnarray}
The $k$ dependence of  $\lambda_F^m,\;\lambda_F^\rho,\;\lambda_F^s$ is obtained from the one-loop expressions \eqref{simpleloop} by adding the infrared cutoff $R_k^F$ to the inverse fermionic propagator and by applying the formal derivative $\tilde\partial_k=(\partial_kR_k^F)\partial/\partial R_k^F$ under the summation. For $\lambda_F^m$, for example, one obtains
\begin{equation}
\partial_k\lambda_F^m(Q)=\tilde\partial_k\Delta\lambda_F^m(Q)\,, \label{looptoflow}
\end{equation}
where the formal derivative $\tilde\partial_k$ should be read as acting under the loop summation of terms contributing to $\Delta\lambda_F^m(Q)$. Note that $(\Delta\lambda_F^m)^F(Q)$ is only part of the complete loop contribution $\Delta\lambda_F^m(Q)$, namely, the one which arises from the two diagrams shown in the first line of Fig. \ref{ff}. The complete $\Delta\lambda_F^m(Q)$ is obtained if the diagrams shown in Fig. \ref{ff} are all taken together.

\begin{figure}[t]
\includegraphics[width=80mm,angle=0.]{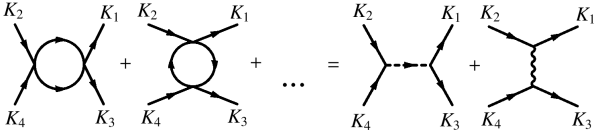}
\caption{\small{Schematic picture of the bosonization of loop contributions to the four-fermion vertex. The terms indicated by the three dots correspond to loop diagrams having internal bosonic lines.}}
\label{bosonization_label}
\end{figure}

In our partially bosonized approach, the fermion loop contributions to the momentum-dependent four-fermion vertex in Eq. \eqref{simpleloop} are fully accounted for by the exchange of the bosons $\mathbf m,\rho$ and $s$. This is shown schematically in Fig. \ref{bosonization_label}. In the language of boson exchange, the momentum dependence of the coupling in, for instance, the magnetic channel can be taken into account by the momentum dependence of the expression $\bar h_m^2(K_1-K_2)\tilde P_m^{-1}(K_1-K_2)$. In practice, we keep $\lambda_F^m=0$ during the flow and account for the loop-generated $\Delta\lambda_F^m$ by a corresponding change $\Delta\bar h_m^2$. We note that only the combination $\bar h_m^2\tilde P_m^{-1}$ appears in the computations as long as the only momentum dependence of the Yukawa couplings is that of the boson momentum. In fact, by a momentum-dependent rescaling of the fields for the $\mathbf m$-boson it is in principle possible to arbitrarily attribute parts of the momentum dependence to $\bar h_m^2$ or to $\tilde P_m$. Nevertheless, introducing the two factors $\bar h_m^2$ and $\tilde P_m^{-1}$ instead of only $\lambda_F^m$ is useful if one wants to approach spontaneous symmetry breaking. It has the advantage that instead of having to deal with the divergent coupling $\lambda_F^m$, one only needs to account for the mass term changing its sign. The term containing $\bar m_m^2$ in Eq. \eqref{m_masse}, which is quadratic in the boson field, becomes part of the effective potential for the magnetic boson in a more extended truncation. Our description of the $k$-dependent flow of $\lambda_F^m$ by means of the $k$-dependent quantities $\bar h_m$ and $\tilde P_m$ (and analogously for $\lambda_F^\rho$ and $\lambda_F^s$) is achieved formally by a $k$-dependent nonlinear field redefinition, see Appendix A, Eq. \eqref{newfield} . At momentum $Q=0$, for example, the contribution to the flow of the momentum-dependent Yukawa couplings due to the diagrams in the first line of Fig. \ref{ff}, according to Eqs. \eqref{rebosrho}, \eqref{rebosm}, is given by
\begin{eqnarray}
\left(\partial_k \bar h_{m/\rho}^2(0)\right)^F&=&-\frac{U^2}{2}\tilde P_{m/\rho}(0)\sum_P\tilde\partial_k\frac{1}{P_F(P)P_F(P)}\,,\label{highscales1}\\
\left(\partial_k \bar h_{s}^2(0)\right)^F&=&\frac{U^2}{2}\tilde P_{s}(0)\sum_P\tilde\partial_k\frac{1}{P_F(P)P_F(-P)}\,.\label{highscales2}
\end{eqnarray}

At this level, we have described the exact one-loop perturbative result for the momentum-dependent four-fermion vertex in terms of boson exchange. The concept of the flowing action, however, allows for a ``renormalization group improvement'' which is obtained by $k$-dependent ``running couplings'' or vertices. In the purely fermionic flows \cite{zanchi1,zanchi2,halbothmetzner,halbothmetzner2,honisalmi,salmhofer,honerkamp01,katanin} the constant coupling $U$ would be replaced by the full momentum- and $k$-dependent four-fermion vertex. In our partially bosonized approach, where we keep only a constant four-fermion coupling $U$, this renormalization group improvement is generated by the diagrams involving internal bosonic lines, shown in Fig. \ref{yukdir} and the second and third lines of \ref{ff}. It is at this level where our truncation for the momentum dependence of the Yukawa couplings and inverse boson propagators as well as the restriction to a certain number of bosons starts to matter.

The momentum dependence of the four-fermion vertex which is generated by boson exchange is much more complicated than the simple form \eqref{simpleloop}. We therefore have to decide how to distribute these contributions onto the different boson exchange channels. To this end, we adopt an approximation where the momentum-dependence of the four-fermion couplings $\lambda_F^m,\;\lambda_F^\rho,\;\lambda_F^s,\;\lambda_F^d$ can be identified with the dependence of the diagrams in Figs. \ref{yukdir} and \ref{ff} on the so-called transfer momentum. This momentum is defined as the difference between the momenta attached to the two fermionic propagators in each diagram. Particle-hole diagrams are absorbed by the real bosons and particle-particle diagrams by the complex Cooper pair bosons. All diagrams are evaluated at external momenta $L=(\pi T,\pi,0)$ and $L'=(\pi T,0,\pi)$ and transfer momenta $0=(0,0,0)$ and $\Pi=(0,\pi,\pi)$. For small values of $|\mu|$ and $|t'|$, the (spatial parts of) momenta $L$ and $L'$ are close to the Fermi surface and the density of states is rather large there, so that this choice will capture the relevant physics for not too large $|\mu|$ and $|t'|$. Where more than one combination of external momenta $\pm L$ and $\pm L'$ is compatible with the condition that the transfer momentum is either $0$ or $\Pi$, we take the average over them. For the coupling in the $d$-wave channel, the evaluation of the contributing diagrams is discussed in more detail in the next section.

While the contributions to the Yukawa couplings in Eqs. \eqref{highscales1} and \eqref{highscales2} are proportional to $U^2$ and therefore present already for large $k$, the diagrams shown in Fig. \ref{yukdir} and in the second and third lines of Fig. \ref{ff} start to have an influence on the flow of the Yukawa couplings only after nonzero Yukawa couplings have been generated due to Eqs. \eqref{highscales1} and \eqref{highscales2} in the first place. In perturbation theory, they would correspond to higher order effects $\sim U^3$ and $U^4$. (Perturbatively, every Yukawa coupling counts as $U$.) The flow of the couplings in the magnetic and charge density channels starts to differ as soon as the diagrams shown in the first line of Fig. \ref{yukdir} become important. They contribute positively to the coupling in the magnetic channel but negatively to the couplings in the charge density and superconducting $s$-wave channels. This explains why among the three Yukawa couplings $\bar h_m\,,\bar h_\rho\,,\bar h_s$ the dominating one is $\bar h_m$, although in accordance with Eqs. \eqref{highscales1} and \eqref{highscales2} all three are generated with equal size at early stages of the flow. Due to the comparatively large Yukawa coupling $\bar h_m$ the mass term $\bar m_m^2$ is driven fastest toward zero by the diagrams in Fig. \ref{bosepropkorr}. We can therefore understand why the charge density and $s$-wave superconducting channels never become critical in the range of parameters investigated.

\section{Coupling in the $d$-wave channel}

The generation of a coupling in the $d$-wave channel in the framework used here has already been discussed in an earlier work.\cite{krahlmuellerwetterich} The $d$-wave Yukawa coupling $\bar h_d$ arises during the renormalization flow due to the first diagram in the lower line of Fig. \ref{ff}, which is the only particle-particle box graph. The coupling is extracted from contributions due to this graph by means of the prescription
\begin{eqnarray}\label{eq:flowlamnda_d}
\Delta\lambda_F^d(\mathbf l,\mathbf l')
&=&\frac{1}{2}\big\{\Delta\Gamma^{(4),pp}_{F,s}(L,L,-L,-L)
\\
& &\hspace{0.4cm}
-\Delta\Gamma^{(4),pp}_{F,s}(L,L',-L,-L')\big\}\nonumber\,,
\end{eqnarray}
where the subscript $s$ denotes the singlet and the superscript $pp$ the particle-particle part of the four-point vertex. The momentum vectors $L$ and $L'$ are defined as in the previous section. For a motivation of this definition of the $d$-wave coupling see \cite{krahlmuellerwetterich}. The contribution from the particle-particle box diagram to the $s$-wave superconducting channel is obtained by adding, instead of subtracting, the two terms on the right-hand side of Eq. \eqref{eq:flowlamnda_d}. The $s$- and $d$-wave superconducting channels of the four-fermion coupling can be described as those parts of its singlet particle-particle contribution which are symmetric ($s$-wave) and antisymmetric ($d$-wave) under a rotation by $90^\circ$ of the outgoing electrons with respect to the incoming electrons. In our approximation, the first diagram in the second line of Fig. \ref{ff} contributes only to the $s$-wave channel.

Once a coupling in the $d$-wave channel has been generated through the particle-particle box diagram, it is further enhanced due to the direct contribution shown as the first graph in the second line of Fig. \ref{yukdir}. Since this graph, which is itself proportional to the Yukawa coupling $\bar h_d$, contributes positively to the flow of $\bar h_d$, it can lead to a growth of this coupling without bounds, i.e. lead to an an instability in the $d$-wave channel. This instability will be the result of antiferromagnetic spin fluctuations (corresponding to the wiggly internal line of the diagram mentioned), so that our results finding a $d$-wave instability through this contribution support the idea, proposed and defended in \cite{miyake,loh,bickers,lee,millis,monthoux,scalapino}, that antiferromagnetic spin fluctuations are responsible for $d$-wave superconductivity in the two-dimensional Hubbard model (and maybe also in the cuprates insofar as the Hubbard model serves as a guide to the relevant cuprate physics).

That the particle-particle graph in the second line of Fig. \ref{yukdir} is crucial for the emergence of a $d$-wave instability arising from antiferromagnetic fluctuations is mirrored by the fact that this diagram has the same momentum structure as the BCS gap equation. In the presence of an interaction which in momentum space is maximal around the  $(\pi,\pi)$-points---a condition which is fulfilled when antiferromagnetic spin fluctuations dominate---the gap solving this equation exhibits $d$-wave symmetry.

\section{Numerical results}

We now come to the discussion of the numerical results we have obtained at small next-to-nearest-neighbor hopping $|t'|/t\leq0.1$. For values of $|t'|$ and $|\mu|$ which are larger than those for which we show results in Fig. \ref{phasediag}, there is, in addition to the tendency to antiferromagnetism which is present already at large scales $k\gg0$, a tendency toward ferromagnetism which becomes important at lower scales. Due to our simple parametrization of the inverse magnetic propagator $\tilde P_m$, see Eqs. \eqref{eq:apropparam1}\,-\,\eqref{eq:apropparaminkomm} in Appendix B, and due to our choice of external momenta for evaluating the diagrams shown in Figs. \ref{yukdir} and \ref{ff}, we may overestimate magnetic fluctuations at larger values of $|\mu|$ and $|t'|$. Hence, we do not show any results for these larger values.

In the range of parameters investigated, we find that either antiferromagnetism or $d$-wave superconductivity is the leading instability. In agreement with previous findings, the coupling in the $d$-wave channel emerges due to antiferromagnetic fluctuations. In the parameter regime where this coupling is enhanced most strongly it competes with (and is driven by) the coupling in the \textit{incommensurate} antiferromagnetic (iAF) channel which was studied in detail in Ref. \onlinecite{simon}, where the same framework was used as here.

\begin{figure}[t]
\includegraphics[width=70mm,angle=0.]{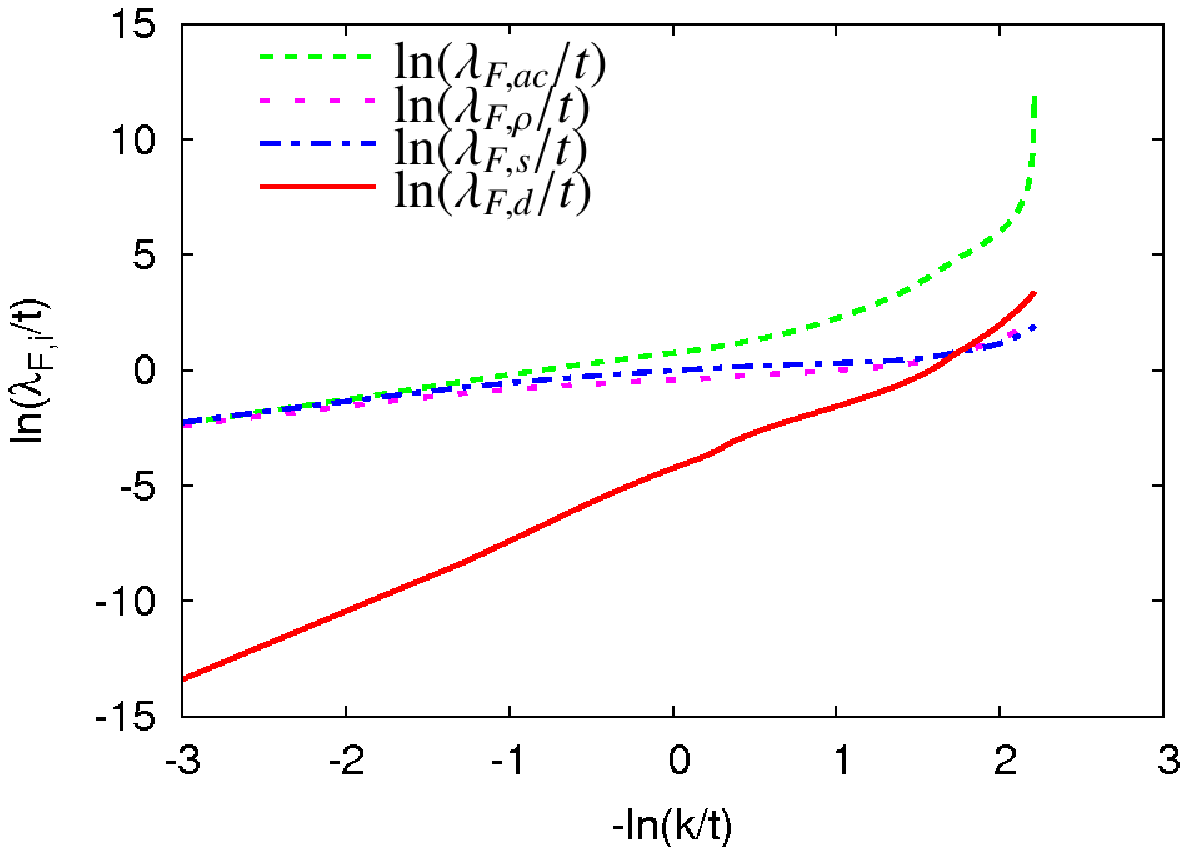}
\includegraphics[width=70mm,angle=0.]{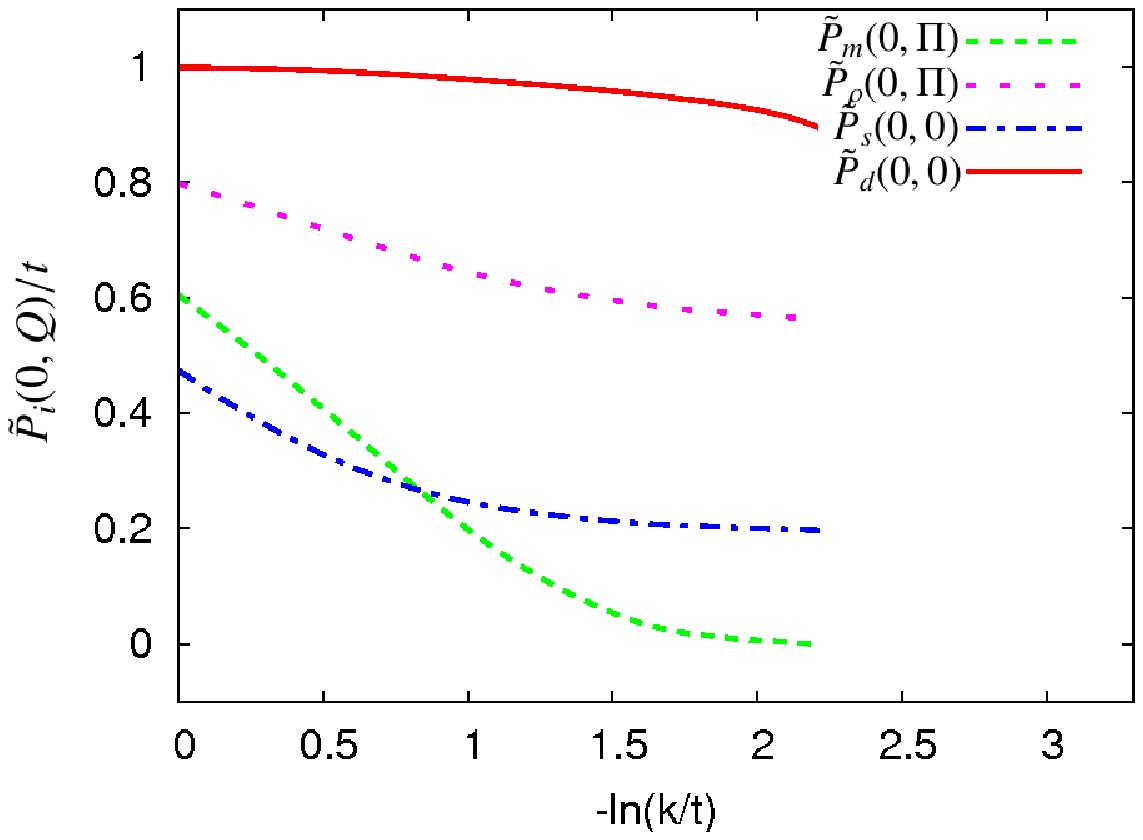}
\caption{\small{Upper panel: flow of the four-fermion vertex in the different channels for $U/t=3$, $t'/t=-0.01$, $\mu/t=-0.12$ and $T/t=0.188$. The shorthands used in the legend are defined as $\lambda_{F,ac}\equiv\bar h_m^2(\Pi)/\tilde P_{m}(\Pi)$, $\lambda_{F,\rho}\equiv\bar h_\rho^2(\Pi)/\tilde P_{\rho}(\Pi)$, $\lambda_{F,s}\equiv\bar h_s^2(0)/\tilde P_s(0)$ and $\lambda_{F,d}\equiv\bar h_d^2(0)/\tilde P_d(0)$. Lower panel: flow of the minima of the inverse bosonic propagators (bosonic mass terms).}}
\label{AFdom}
\end{figure}

\begin{figure}[t]
\includegraphics[width=70mm,angle=0.]{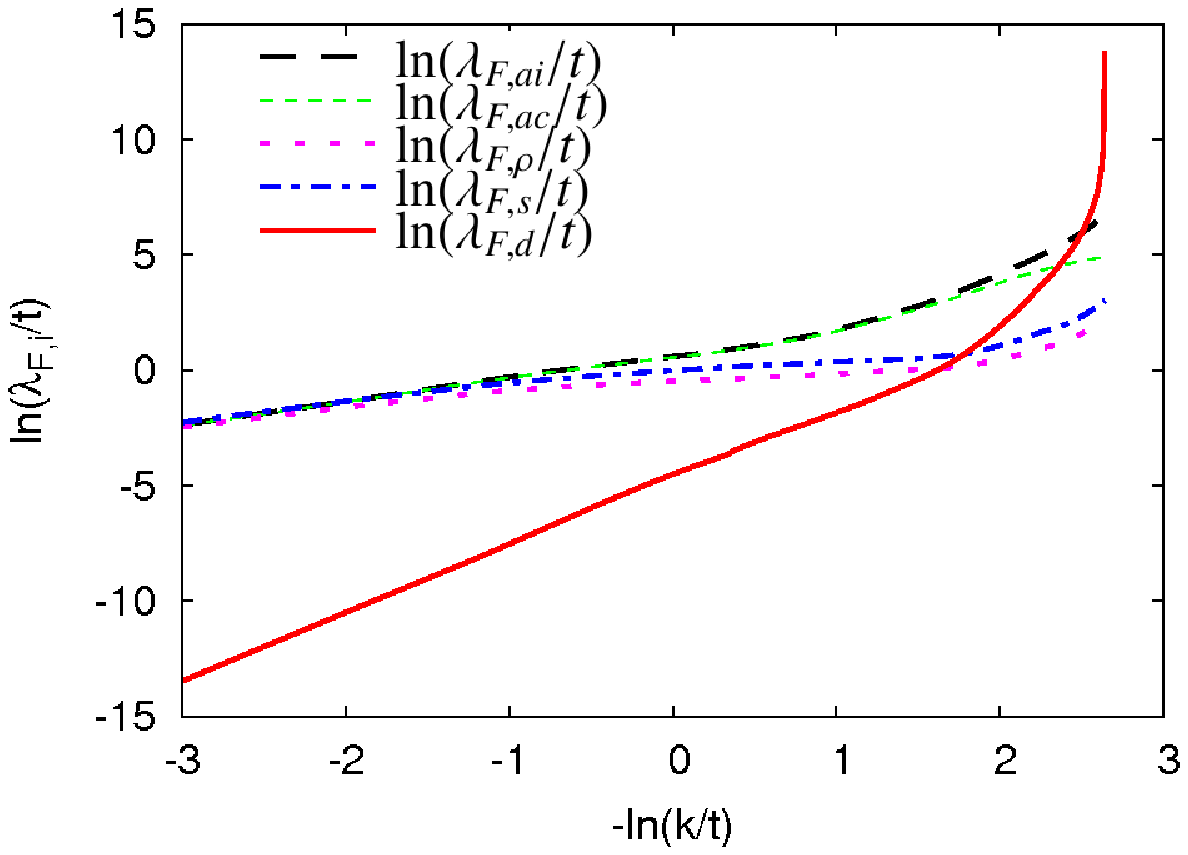}
\includegraphics[width=70mm,angle=0.]{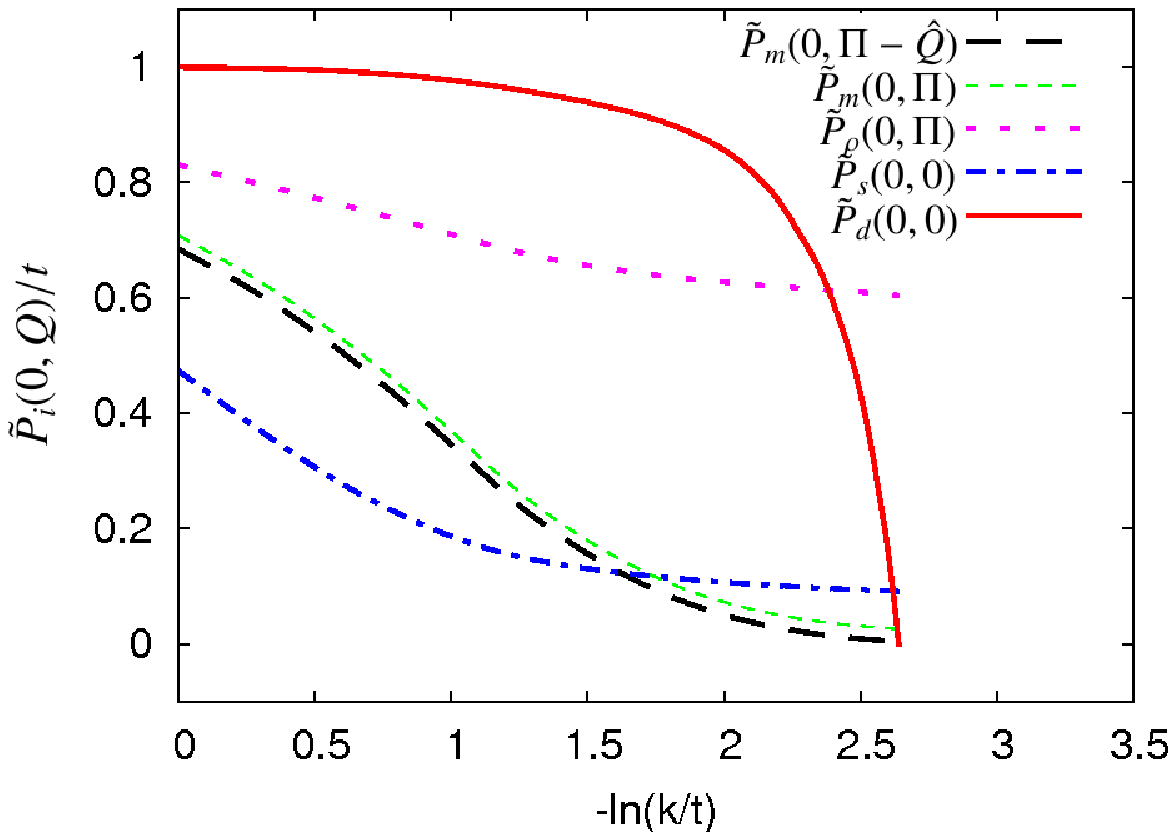}
\caption{\small{Same as Fig. \ref{AFdom} for $\mu/t=-0.32$ and $T/t=0.109$. In addition to the couplings defined in Fig \ref{AFdom} we plot $\lambda_{F,ai}\equiv\bar h_m^2(\Pi-\hat Q)/\tilde P_{m}(\Pi-\hat Q)$ (coupling in the incommensurate AF channel), where $\hat Q=(0,\hat q,0)$ with $\hat q$ the size of the incommensurability.  For these parameters the coupling in the $d$-wave channel diverges first. In the magnetic channel, incommensurate antiferromagnetic fluctuations (long-dashed lines) dominate over commensurate ones (short-dashed lines).}}
\label{SCdom}
\end{figure}

\begin{figure}[t]
\includegraphics[width=70mm,angle=0.]{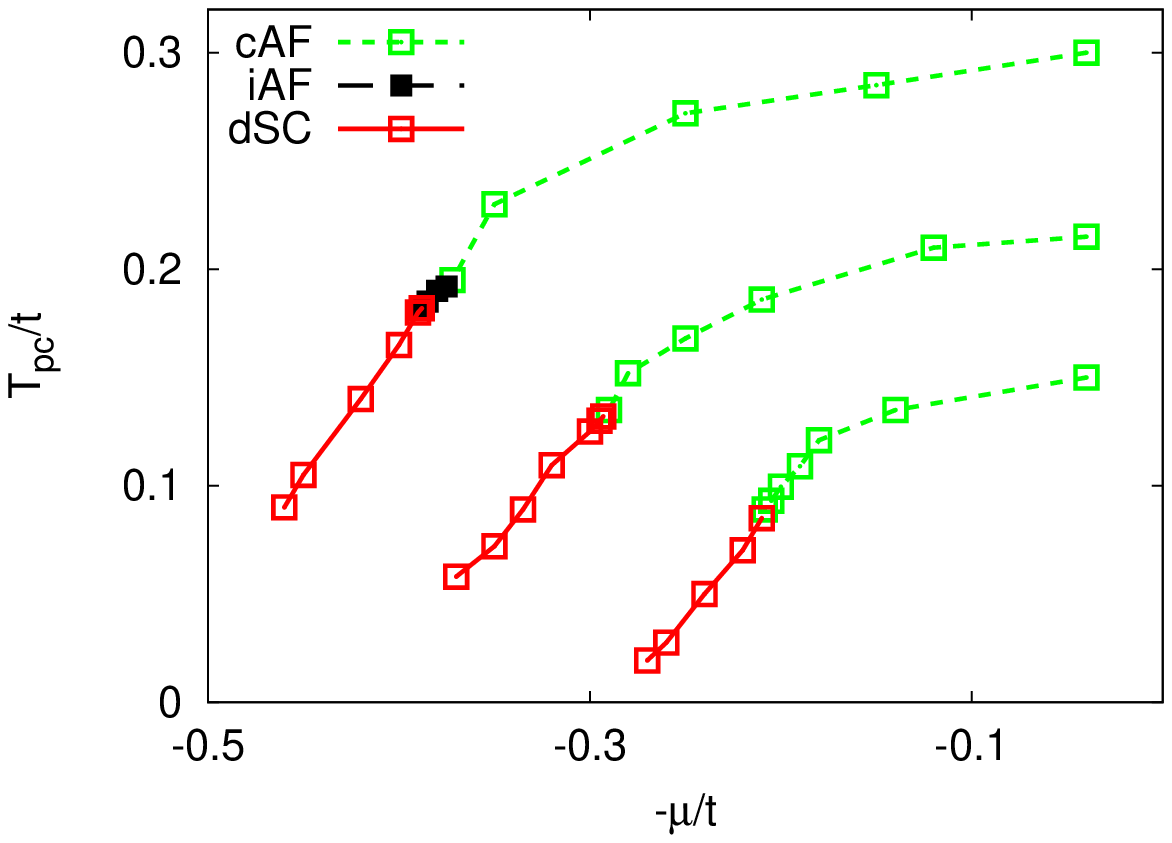}
\includegraphics[width=70mm,angle=0.]{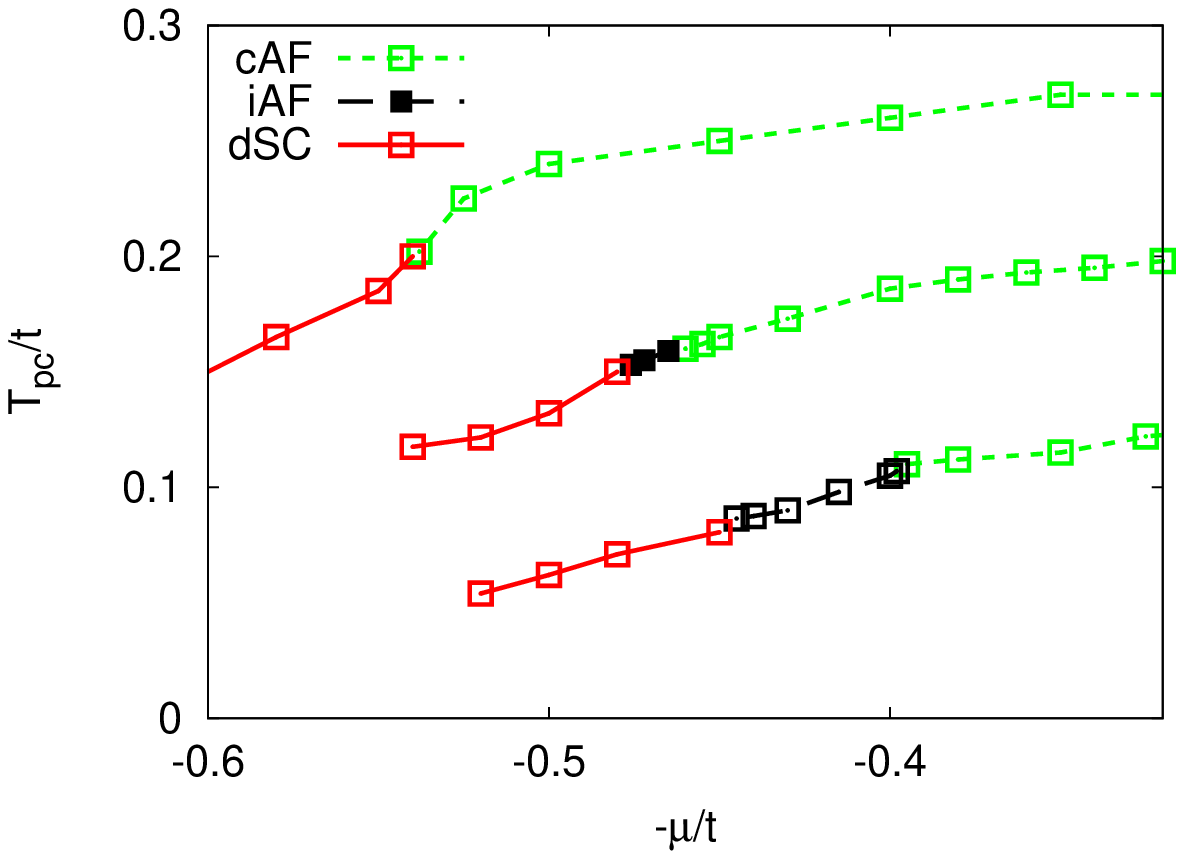}
\caption{\small{Pseudocritical temperatures $T_{pc}$ for $t'/t=-0.01$ (upper panel) and $t'/t=-0.1$ (lower panel) at different values of $U/t=3.5$ (upper lines), $U/t=3$ (middle lines) and $U/t=2.5$ (lower lines). Short-dashed lines denote the onset of commensurate antiferromagnetic order (cAF), long-dashed lines, appearing in small regions at larger values of $-\mu$, the onset of incommensurate antiferromagnetic order (iAF). Solid lines indicate the onset of $d$-wave superconducting order (dSC)).}}
\label{phasediag}
\end{figure}

In the upper panels of Figs. \ref{AFdom} and \ref{SCdom}, the flow of the different channels of the fermionic four-point vertex is displayed at fixed $t'/t=-0.01$ and different values of $U/t=2.5\,,3\,,3.5$. The lower panels show the flow of the corresponding bosonic masses, which approximate unrenormalized inverse susceptibilities in channels which are close to critical. As expected, the antiferromagnetic coupling grows fastest and remains the dominant one for small to intermediate values of $|\mu|$, for a representative case see Fig. \ref{AFdom}. For $\mu/t<-0.28$, however, the $d$-wave coupling diverges for higher temperatures than the antiferromagnetic coupling, for an example of this kind of scenario see Fig. \ref{SCdom}. The couplings in the charge density wave and superconducting $s$-wave channels are also considerably enhanced in both cases, and their influence is quantitatively important although they do not diverge.

In Fig. \ref{phasediag} the highest temperature at a given value of $\mu$ is plotted for which one of the boson masses drops to zero at some scale $\bar k$, signaling the onset of local order on a typical length scale $\bar k^{-1}$ in the corresponding channel. These ``pseudocritical temperatures'' $T_{pc}$ are shown for $t'/t=-0.01$ (upper panel) and $t'/t=-0.1$ (middle panel) and different values of $U$. Pseudocritical temperatures for antiferromagnetism are higher by a factor $\approx 3$ than those presented in our last paper.\cite{simon} This is mainly due to the neglect of the fermionic wave function renormalization and of quartic bosonic couplings in the present paper. Both of these would suppress the growth of the four-fermion vertex and hence the emergence of local order. These contributions are omitted here for the sake of a simple and nevertheless systematic approach to the four-fermion vertex. They will be included in a forthcoming work where also the bosonic vertex functions that directly couple together the different types of bosons will be taken into consideration. We recall that often the true critical temperature $T_c$ is found to be substantially smaller than the pseudocritical temperature $T_{pc}$.\cite{bbw04,kw07}

For some pairs of parameters $U$ and $t'$ there exists a range of values of $\mu$ where incommensurate antiferromagnetism has the largest pseudocritical temperature, for others there are no such values of $\mu$, see Fig. \ref{phasediag}. In the range of $\mu$, where the transition from (either commensurate or incommensurate) antiferromagnetic to $d$-wave superconducting order occurs in Fig. \ref{phasediag}, there is an extremely close competition between the couplings in the commensurate and incommensurate antiferromagnetic and $d$-wave superconducting channels. Which of them diverges first may in part depend on the truncation, so when one includes fermionic self-energy contributions and quartic bosonic couplings, this may have an important effect on the size and existence of regions exhibiting local incommensurate antiferromagnetic order.

Most existing studies using the framework of the fermionic functional renormalization group focus on the flow of the four-fermion vertex, as we do in the present work, so we can compare our results with theirs. The most recent results for the phase diagram of the two-dimensional Hubbard model at varying $\mu$ and fixed $t'$, presented in \cite{andrei}, are obtained by means of an $N$-patch scheme restricted to the flow of the four-fermion vertex. A temperature-flow scheme is used in that work, where the temperature is used as a parameter which flows from infinity to a nonzero value where a first vertex function reaches some critical value. The temperature $T^*$ where the divergence of the four-fermion vertex occurs (it is obtained in \cite{andrei} by means of a polynomial fit of the inverse susceptibilities) may be compared to our pseudocritical temperature $T_{pc}$. Both correspond, albeit in different ways, to the divergence of the four-fermion vertex and the onset of local order. However, whereas in the temperature-flow scheme the temperature appears as a flow parameter, it is kept fixed in our approach.

The results in \cite{andrei} are in complete \textit{qualitative} agreement with ours. This concerns, for instance, the dependence of local order on the values of $U$ and $t'$. If $U$ is increased, the divergence of vertex functions (equivalent to the emergence of local order) occurs already at larger temperatures and the divergence of the coupling in the $d$-wave channel, for the small values of $|t'|$ considered, happens closer to the van Hove filling $\mu=4t'$. There is also agreement on the fact that $d$-wave superconductivity, when its coupling is enhanced most strongly, competes mostly with incommensurate rather than commensurate antiferromagnetism (cAF) as the dominant instability.

The quantitative comparison between our results and those in \cite{andrei} has to be handled with some care: for $T\leq T_{pc}$ our flow is stopped at a nonzero scale $k$. For quantities evaluated at $k\neq0$, the detailed implementation of the infrared cutoff has an effect on the results. (This contrasts with results for temperatures above the pseudocritical line in Fig. \ref{phasediag}, where the four-fermion vertex never diverges, such that we can extrapolate to $k=0$. For $k=0$, any residual dependence on the cutoff scheme is an indication of the shortcomings of a given truncation.) Despite this caveat, the comparison remains instructive. We find that for $t'/t=-0.1$ and $U/t=3.5$ the maximal values of $T_{pc}$ and  $T^*$ as functions of $\mu$ differ by a factor of about $4/3$, and slightly more for $U/t=2.5$. For the onset of $d$-wave superconducting order, the pseudocritical temperature $T_{pc}$ as a function of $\mu$ shown in Fig. \ref{phasediag} is larger than $T^*_{dSC}$ obtained in \cite{andrei} by a factor of at least $2$ for both $U/t=2.5$ and $U/t=3.5$. The difference gets more pronounced with increasing distance from half filling. As already mentioned, a possible source of quantitative shortcomings of our calculations is that magnetic fluctuations may be overestimated due to our simple parametrization of the momentum dependence of the magnetic boson propagator (see Eqs. \eqref{eq:apropparam1}\,-\,\eqref{eq:apropparaminkomm}). Furthermore, the diagrams shown in Figs. \ref{yukdir} and \ref{ff} are evaluated at $(0,\pi)$ and $(\pi,0)$, which is adequate only for not too large values of $|\mu|$ where the Fermi surface is close to the boundary of the Brillouin zone. When magnetic fluctuations are generally overestimated and antiferromagnetic fluctuations are dominant, the critical scales for the onset of $d$-wave superconducting order and hence the pseudocritical temperatures can be expected to come out too large. We recall at this place that both the present work and Ref.\ \onlinecite{andrei} neglect the renormalization of the fermionic propagator, which is expected to have a sizeable lowering effect on the value of $T_{pc}$.


\section{Conclusion}

In this work we have shown that the functional renormalization group approach to correlated fermion systems based on partial bosonization can account for the competition between the antiferromagnetic and superconducting instabilities in the two-dimensional Hubbard model. We have studied the emergence of a coupling in the $d$-wave channel and its divergence in a certain parameter range as a consequence of antiferromagnetic spin fluctuations. In a nutshell, this result confirms the spin-fluctuation route to $d$-wave superconductivity in the two-dimensional Hubbard model.

Our treatment of the fermionic four-point vertex paves the way for a unified treatment of spontaneous symmetry breaking in the two-dimensional Hubbard model. In a next step, self-energy corrections to the electrons as well as quartic bosonic couplings can be included in our approach. This may shed light on the question of coexistence of different types of order, which so far has been addressed in the framework of the functional renormalization group only on the basis of a mean field approach replacing the flow of vertex functions at lower scales.\cite{metznerreissrohe,reiss} In a final step, a unified treatment of the flow of vertex functions in both the symmetric and symmetry-broken regimes can be given within the present approach.

{\bf Acknowledgments}: We are grateful to C. Husemann, A. Katanin and M. Salmhofer for useful discussions. SF acknowledges support by Studienstiftung des Deutschen Volkes.

\appendix

\section{Flowing bosonization}
To illustrate how flowing bosonization works, consider, as an example, the part of the effective average action which in our truncation involves the $\mathbf m$-boson. (The computation is exactly analogous for the other bosons.) It is given by
\begin{eqnarray}
&&\Gamma_{m}+\Gamma_{Fm}+\Gamma_F^m=\\
&&\hspace{1cm}\frac{1}{2}\sum_{Q}\mathbf m^T(-Q)\left(P_{m}(Q)+m_m^2\right)\mathbf m(Q)\nonumber\\
&&\hspace{1cm}+\sum_{K,Q,Q'}\bar h_m(K)\;\mathbf m(K)\cdot[\psi^{\dagger}(Q)\boldsymbol\sigma\psi(Q')]\;\delta(K-Q+Q')\nonumber\\
&&\hspace{1cm}-\frac{1}{2}\sum_{K_1,K_2,K_3,K_4}\lambda_F^m(K_1-K_2)\delta\left( K_1-K_2+K_3-K_4 \right)\nonumber\\
&&\hspace{1.5cm}\times\,\big\lbrack \psi^\dagger(K_1)\boldsymbol\sigma\psi(K_2) \big\rbrack\cdot\big\lbrack \psi^\dagger(K_3)\boldsymbol\sigma\psi(K_4) \big\rbrack\nonumber\,.
\end{eqnarray}
Now we introduce a scale dependence of the field $\mathbf m(Q)$, writing it as $\mathbf m_k(Q)$. The change in $\mathbf m_k(Q)$ between two scales $k$ and $k-\Delta k$ that are infinitesimally close to each other can be chosen to be
\begin{eqnarray}
\mathbf m_k(Q)-\mathbf m_{k-\Delta k}(Q)=\Delta\alpha_k(Q)\mathbf{\tilde m}_k(Q)\,,
\end{eqnarray}
where the auxiliary field $\mathbf{\tilde m}_k(Q)$ is given by
\begin{equation}
\mathbf{\tilde m}_k(Q)=\sum_P[\psi^{\dagger}(P)\boldsymbol\sigma\psi(P-Q)]\,,
\end{equation}
and $\alpha_k(Q)$ is a function that will be chosen in such a way that $\lambda_F^m$ cancels to zero at all scales.

To achieve this, we take the generalized flow equation
\begin{eqnarray}
\partial_k\Gamma[\chi_k]=\partial_k\Gamma[\chi_k]\big|_{\chi_k}+\sum_Q\left(\partial_k\chi_k\right)\frac{\delta\Gamma_k[\chi_k]}{\delta\chi_k}\,,
\end{eqnarray}
which yields in our case
\begin{eqnarray}
\partial_k\Gamma_k=\partial_k\Gamma_k\big|_{\mathbf m_k}+\sum_Q\left( -\partial_k\alpha_k(Q)\tilde P_{m,k}(Q)\mathbf m_k(Q)\cdot\mathbf{\tilde m}_k(Q)\right.\nonumber\\
\left.+\partial_k\alpha_k(Q)\bar h_m(Q)\mathbf{\tilde m}_k(Q)\cdot\mathbf{\tilde m}_k(-Q) \right)\,.\label{newfield}
\end{eqnarray}
We can read off the modified equations for $\lambda_F^m$ and $\bar h_m$ and set the scale-dependence of $\lambda_F^m$ to zero,
\begin{eqnarray}
\partial_k\bar h_m(Q)&=&\partial_k\bar h_m\big|_{m_k}(Q)+\tilde P_{m,k}(Q)\partial_k\alpha_k(Q)\,,\\
\partial_k\lambda_F^m(Q)&=&\partial_k\lambda_F^m\big|_{\mathbf m_k}(Q)+2\bar h_m(Q)\partial_k\alpha_k(Q)\equiv0\,.\nonumber
\end{eqnarray}
This allows us to eliminate the hitherto undetermined function $\alpha_k(Q)$ and to obtain the flow equation for the Yukawa coupling including contributions from flowing bosonization,
\begin{eqnarray}
\partial_k\bar h_m(Q)&=&\partial_k\bar h_m\big|_{\mathbf m_k}(Q)+\frac{\tilde P_{m,k}(Q)}{2\bar h_m(Q)}\partial_k\lambda_F^m\big|_{\mathbf m_k}(Q)\,.\label{rebosrho}
\end{eqnarray}
Similarly, we get for the other Yukawa couplings the following contributions from flowing bosonization:
\begin{eqnarray}
\partial_k\bar h_\rho(Q)&=&\partial_k\bar h_\rho\big|_{\rho_k}(Q)+\frac{\tilde P_{\rho,k}(Q)}{2\bar h_\rho(Q)}\partial_k\lambda_F^\rho\big|_{\rho_k}(Q)\,,\nonumber\\
\partial_k\bar h_s(Q)&=&\partial_k\bar h_s\big|_{s_k,s^*_k}(Q)+\frac{\tilde P_{s,k}(Q)}{2\bar h_s(Q)}\partial_k\lambda_F^s\big|_{s_k,s^*_k}(Q)\,,\label{rebosm}\\
\partial_k\bar h_d(Q)&=&\partial_k\bar h_d\big|_{d_k,d^*_k}(Q)+\frac{\tilde P_{d,k}(Q)}{2\bar h_d(Q)}\partial_k\lambda_F^d\big|_{d_k,d^*_k}(Q)\,.\nonumber
\end{eqnarray}

\section{Parametrization of bosonic propagators and Yukawa couplings}
In our truncation, both the $\tilde P_i$ and $\bar h_i$ are momentum-dependent functions which in addition depend on the scale $k$. In principle, one could try to discretize the momentum dependence and attempt a numerical solution of the partial differential equations for $\tilde P_i(Q,k)$ and $\bar h_i(Q,k)$. Instead, we proceed in this paper to a parametrization of the momentum dependence which we describe in this appendix.

\subsection{Propagators}
The truncation of the inverse bosonic propagators is briefly described in the following lines. For a more detailed discussion see \cite{simon}. (Note, however, that in \cite{simon} we discuss the antiferromagnetic propagator which is distinguished from the magnetic propagator employed here by a shift in the argument by the antiferromagnetic wave vector $\boldsymbol\pi=(\pi,\pi)$.) For the kinetic term $P_m$ of the magnetic boson we make the ansatz
\begin{eqnarray}\label{eq:apropparam}
P_{m,k}(Q)=Z_m\omega_Q^2+A_mF(\mathbf q)\,.
\end{eqnarray}
The quadratic dependence on frequency is motivated by mean field results for small $|\omega_Q|$. (For larger values of $|\omega_Q|$, it mimics the decaying frequency-dependence of the Yukawa couplings, which is not taken into account explicitly.)

In Eq. \eqref{eq:apropparam} we employ for $F(\mathbf q)$
\begin{eqnarray}\label{eq:apropparam1}
F_c(\mathbf q)=\frac{D_m^2\cdot[\mathbf{q}-\boldsymbol\pi]^2}{D_m^2+[\mathbf{q}-\boldsymbol\pi]^2}
\,,\end{eqnarray}
if commensurate antiferromagnetic fluctuations dominate. Here $[\mathbf{q}]^2$ is defined as $[\mathbf{q}]^2=q_x^2+q_y^2$ for $q_{x,y}\in [-\pi,\pi]$ and continued periodically otherwise. If incommensurate antiferromagnetic fluctuations dominate, we use
\begin{eqnarray}\label{eq:apropparam2}
F_i(\mathbf q,\hat q)&=&\frac{D_m^2\tilde F(\mathbf q,\hat q)}{D_m^2+\tilde F(\mathbf q,\hat q)}
\,,\end{eqnarray}
where the momentum dependence is quartic in momentum and explicitly includes the incommensurability $\hat q$,
\begin{eqnarray}\label{eq:apropparaminkomm}
\tilde F(\mathbf q,\hat q)=\frac{1}{4\hat q^2}\big((\hat q^2-[\mathbf q-\boldsymbol\pi]^2)^2+4[q_x-\pi]^2[q_y-\pi]^2\big)
\,.\end{eqnarray}

The shape coefficient $D_m$ used in Eqs. \eqref{eq:apropparam1}, \eqref{eq:apropparam2} is computed as
\begin{eqnarray}\label{eq:D}
D_m^2=\frac{1}{A_m}\big(P_m(0,0,0)-P_m(0,\pi-\hat q,\pi)\big).
\end{eqnarray}
The $Z_m$- and $A_m$-factors are computed from the differences of inverse propagators at different frequencies and momenta,
\begin{eqnarray}\label{eq:AZ}
Z_m&=&\frac{1}{\left(2\pi T\right)^2}\,\left(P_{m,k}(2\pi T,\mathbf q=0) -P_{m,k}(0,\mathbf q=0) \right)\,,\nonumber\\
A_m&=&\frac{1}{\overline q^2}\,\left(P_{m,k}(0,\hat q+\overline q, 0) - P_{m,k}(0,\hat q, 0) \right)\,,
\end{eqnarray}
where $\overline q$ is a parameter which is fixed in such a way that results are practically independent of it. For the results displayed before, we have set it to $0.15$. The propagator of the $\rho$-boson is treated in exactly the same way.

For the $s$-and $d$-bosons, the treatment is just as for the $a$- and $\rho$-bosons in the commensurate case. Since the minima of the inverse propagator do not occur in the vicinity of the wave vector $\boldsymbol\pi$, we do not use a shift by this vector,
\begin{eqnarray}
P_{s/d,k}(Q)=Z_{s/d}\omega^2+A_{s/d}F_{s/d}(\mathbf q)
\end{eqnarray}
with
\begin{eqnarray}
F_{s/d}(\mathbf q)=\frac{D_{s/d}^2\cdot[\mathbf{q}]^2}{D_{s/d}^2+[\mathbf{q}]^2}\,.
\end{eqnarray}

\subsection{Yukawa couplings}
 The Yukawa couplings $\bar h_m(Q)$ and $\bar h_\rho(Q)$ are parametrized by means of a linear momentum dependence
\begin{eqnarray}
\bar h_{m/\rho}(Q)=\frac{|\boldsymbol\pi-\mathbf q|}{|\boldsymbol\pi|}\bar h_{m/\rho}(0)+\frac{|\mathbf q|}{|\boldsymbol\pi|}\bar h_{m/\rho}(\Pi)\,,
\end{eqnarray}
and the contributions to the flow of $\bar h_{m/\rho}(0)$ and $\bar h_{m/\rho}(\Pi)$ are computed according to Eqs. \eqref{rebosrho} and \eqref{rebosm},
\begin{eqnarray}
\partial_k \bar h_{m/\rho,k}^2(0)&=&\partial_k \bar h_{m/\rho,k}^2\big|_{\mathbf{m}_k/\rho_k}(0)-\tilde P_{m/\rho,k}(0)\partial_k\lambda_{F,k}^{m/\rho}\big|_{\mathbf{m}_k/\rho_k}(0)\,,\nonumber\\
\partial_k \bar h_{m/\rho,k}^2(\Pi)&=&\partial_k \bar h_{m/\rho,k}^2\big|_{\mathbf{m}_k/\rho_k}(\Pi)-\tilde P_{m/\rho,k}(\Pi)\partial_k\lambda_{F,k}^{m/\rho}\big|_{\mathbf{m}_k/\rho_k}(\Pi)\,.\nonumber\\
\label{average0}
\end{eqnarray}
This approximation is most adequate when the loop contributions to $\lambda_F^{m/\rho}(K_1-K_2)$ are minimal for $K_1-K_2=0$ and maximal for $K_1=K_2=\Pi$ or inversely. This is the case whenever either ferromagnetic or \textit{commensurate} antiferromagnetic fluctuations dominate. When \textit{incommensurate} antiferromagnetic fluctuations dominate, we approximate the loop contribution to $\bar h_{m/\rho}(\Pi)$ occurring in the second line of Eq.\ \eqref{average0} by
\begin{eqnarray}
\partial_k \bar h_{m/\rho,k}^2(\Pi)&=&\partial_k \bar h_{m/\rho,k}^2\big|_{\mathbf{m}_k/\rho_k}(\Pi)\\
              && -\tilde P_{m/\rho,k}(\Pi-\hat Q)\,\partial_k\lambda_{F,k}^{m/\rho}\big|_{\mathbf{m}_k/\rho_k}(\Pi)\nonumber\,.
\end{eqnarray}
Here $\hat Q$ denotes an incommensurability the size of which is given by the positions of the minima of the inverse magnetic (or charge density) propagator. The maximal value of $\bar h_m(0,\mathbf q)$, which in the parameter regime we study is always located more closely to $(\pi,\pi)$ than to $(0,0)$, has been denoted $\bar h_a$ in \cite{bbw04,bbw05,krahlmuellerwetterich,simon}. The minimal value of $\tilde P_m(Q)$ corresponds to the ``antiferromagnetic mass term'' $\bar m_a^2$ introduced there.

For the $s$- and $d$-bosons, dominant contributions are for $K_1=-K_3$. This is partly accounted for by the propagators, and for the $s$-boson no further momentum-dependence of the Yukawa coupling is assumed. When the $d$-wave channel becomes critical (see Fig. 7\,(b) of \cite{husemann}), the four-fermion coupling in the $d$-wave channel has a sharp peak around zero momentum. This is accounted for by including a Gaussian function which is centered around zero momentum in the definition of $\bar h_d$. We have checked that our results are practically independent of the width of this Gaussian function, as long as it is reasonably peaked.

\vspace{0.5cm}

\renewcommand{\thesection}{}
\renewcommand{\thesubsection}{A{subsection}}
\renewcommand{\theequation}{A \arabic{equation}}


\end{document}